\documentclass[aps,prd,twocolumn,english]{revtex4-2}
\usepackage[T1]{fontenc}
\usepackage[utf8]{luainputenc}
\usepackage{amsmath,amsfonts,amsthm,bm}
\usepackage{braket}
\usepackage{amssymb}
\usepackage{graphicx}
\usepackage{esint}
\usepackage{babel}

\begin{document}

\title{Tomita-Takesaki theory and quantum concurrence}
	\author{Rupak Chatterjee}
 	\affiliation{
		Department of Applied Physics, New York University, 2 MetroTech Center, Brooklyn, NY, 11201}
	\email{Rupak.Chatterjee@nyu.edu}
    \date{\today}

	\begin{abstract}
 The quantum entanglement measure of concurrence is shown to be directly calculable from a Tomita- Takesaki modular operator framework constructed from the local von Neumann algebras of observables for two quantum systems. Specifically, the Tomita-Takesaki modular conjugation operator $J$ that links two separate systems with respect to their von Neumann algebras is related to the quantum concurrence $C$ of a pure bi-variate entangled state composed from these systems. This concurrence relation provides a direct physical meaning to $J$ as both a symmetry operator and a quantitative measure of entanglement. This procedure is then demonstrated for a supersymmetric quantum mechanical system and a real scalar field interacting with two entangled spin-$\frac{1}{2}$ Unruh-DeWitt qubit detectors. For the latter system, the concurrence result is shown to be consistent with some known results on the Bell-CHSH inequality for such a system.    
	\end{abstract}
\maketitle

\section{Introduction}

Symmetry transformations have long been a key mathematical technique in physics that often eases the analysis of a system by dramatically simplifying various calculations. 
Tomita-Takesaki symmetry refers to those systems that are linked by their respective von Neumann algebra of observables where the set of observables of one system commute with those of the other. The Tomita-Takesaki modular conjugation operator $J$ maps one von Neumann algebra to the other effectively creating a duality between these separate localized systems. One may create an entangled state between these dual systems and try to quantify the degree of entanglement between them. In two previous works \cite{Chatterjee1}, \cite{ Chatterjee2}, it was shown that the quantitative measure of entanglement called concurrence may be derived directly from the modular conjugation operator $J$ itself thus providing another physical meaning to this symmetry operator as a quantitative measure of entanglement.

Why does one need the machinery of operator algebras and specifically von Neumann algebras (\cite{Witt2022, Witt2018, Sewell2002})?  Consider a simple spin Hamiltonian on $n$-sites
$H = - \sum_{i=1}^{n} \sigma^z_i$
with non-interacting sites and standard Pauli matrix algebra.
If $n$ is finite, the algebra of observables formed from the Pauli spin operators has a unique irreducible representation, whereas if $n$ is infinite, the representation will not be unique. A potential Hilbert space for the infinite case that contains all states of the countable infinite set of spins has an uncountable infinite dimension.  One would rather work with a separable Hilbert sub-space $\mathcal{H}$ such that its dimension is countably infinite.
Consider an algebra of observables $\mathcal{A}_0$ that consists of all finite linear combinations of tensor products over sites of polynomials of single Pauli spin operators that satisfy the Pauli matrix algebra. 
One may begin to construct a separable Hilbert sub-space starting with the ground state of our Hamiltonian, $\Omega_{\uparrow} = \ket{\uparrow} \otimes \ket{\uparrow} \otimes \ket{\uparrow} \otimes \cdots$.
If we wish $\mathcal{A}_0$ to remain our algebra of observables, we must include every state in our Hilbert space that is created by acting with $\mathcal{A}_0$ on our ground state $\Omega_{\uparrow}$. Since spins per site are either spin-up or spin-down, the set of states we get from $\mathcal{A}_0\Omega_{\uparrow}$ will consist of states that where all but finitely many of spin sites will be spin-up since $\mathcal{A}_0$ is made up of finite linear combinations of spin operators. Adding weak operator limits of
elements of $\mathcal{A}_0$ creates an algebra $\mathcal{A}$ of bounded operators on this separable Hilbert space $\mathcal{H}$. A weak operator limit of a sequence of elements of $\mathcal{A}_0$,  $\mathfrak{a}_1,\;\mathfrak{a}_2, \cdots  \in \mathcal{A}_0$ converges if the following limit exists
$\lim_{n \rightarrow \infty}\bra{\psi} \mathfrak{a}_n \ket{\phi} < \infty, \forall \psi, \phi, \in \mathcal{H}$. For example, the spin-raising operator
$\sigma^{+}_i = \dfrac{1}{2}\left(\sigma^{x}_i+i\sigma^{y}_i  \right)$ ,   
has a finite weak operator limit in $\mathcal{H}$,
 $ \lim_{n \rightarrow \infty}\bra{\psi} (\sigma^{+}_n)^{\dagger} \sigma^{+}_n \ket{\phi} =0$
since nearly all states in our Hilbert space have the majority of states with sites being spin-up. The algebra $\mathcal{A}$ constructed with its weak operator limits is a \textit{von Neumann algebra} (see the Appendix for the precise definition).

One didn't need to start with $\Omega_{\uparrow}$ but could have started with all spins down,
$\Omega_{\downarrow} = \ket{\downarrow} \otimes \ket{\downarrow} \otimes \ket{\downarrow} \otimes \cdots$. One could have created an algebra of observables $\mathcal{A}'$ with weak operator limits and a separable Hilbert space $\mathcal{H}'$ as above using $\Omega_{\downarrow}$. The key point here is that $\mathcal{A}'$ \textit{will not be the same as} $\mathcal{A}$. For instance, the weak operator limit of spin-raising operators in this case is
$\lim_{n \rightarrow \infty}\bra{\psi} (\sigma^{+}_n)^{\dagger} \sigma^{+}_n \ket{\phi} =1$,
since nearly all states in our Hilbert space have the majority of states with sites being \textit{spin-down}. Therefore, the von Neumann algebras $\mathcal{A}'$ and $\mathcal{A}$ have different weak limit closures. This is not the case of a finite spin model where all algebras, up to isomorphism, have a unique irreducible representation. Therefore, von Neumann algebras, a special type of $\bm{C}^*$-algebra of bounded operators on a Hilbert space that is closed in the weak operator topology, are needed to fully analyze infinite dimensional systems.

Operator algebras have also been used to formulate a general mathematical structure that can encompass classical mechanics, quantum mechanics, statistical mechanics, and quantum field theory. This algebraic approach is where the  \textit{observables} of a system form a $\bm{C^*}$-algebra $\mathcal{A}$ and the \textit{states} of the system $\bm{\omega}$ provide a probability distribution. That is, the pairing of an element of $\mathcal{A}$ and $\bm{\omega}$ produces an expectation value making  $\bm{\omega}$ a \textit{functional} on $\mathcal{A}$ that may be connected to an experimental result. The \textit{dynamics} of the systems are driven by a one-parameter ($t$) group of $\bm{C^*}\text{-automorphisms} \;\;
\mathcal{U}_t: \mathcal{A} \rightarrow \mathcal{A}$.

In classical physics, the states define a probability distribution in phase space whereas the observables are the commutative real-valued functions on phase space. In quantum physics, the states will be seen as linear maps $\bm{\omega}:  \mathcal{A} \rightarrow \mathbb{C}$. The common Hilbert space structure will be an emergent concept from the primary concepts of states $\bm{\omega}$ and the algebra of observables $\mathcal{A}$.
Classical physics will emerge as a probability theory on commutative algebras where as quantum physics will emerge as a probability theory on non-commutative algebras. The general structure of either system is a triplet $(\mathcal{A}, \{\bm{\omega} \}, \mathcal{U}_t)$ of observables, states, and dynamics.

The paper is organized as follows.  Sec. II reviews several aspects of operator algebras that will be needed in this work. Sec. III discusses the quantum entanglement measure called concurrence and its relation to the Bell-CHSH inequality. This relation will be used to make a comparison of the results in this paper to some other previous work in this area. Sec. IV illustrates the use of the modular conjugation operator $J$ on a supersymmetric quantum mechanical system by constructing two dual systems and measuring the entanglement of a bi-variate state between them.  Sec V. starts by describing a Unruh-DeWitt qubit detector model that interacts with a scalar quantum field. It then calculates the concurrence between an entangled pair of qubit detectors interacting with the quantum field by using the Tomita-Takesaki modular operator framework similar to the one of Sec. IV.  Finally, the result is compared to the Bell-CHSH inequality derived in \cite{Guedes2024}.

\section{Smeared Fields and Operator Algebras}

It is common in the literature to find equal time commutation relations of Bosonic field operators of the form $[\phi(\mathbf{x}), \phi^{\dagger}(\mathbf{y})]=i\delta(\mathbf{x}-\mathbf{y})$. Such a definition is merely formal as no Hilbert space has been defined and the delta function $\delta(\mathbf{x}-\mathbf{y})$, which is not really a function, is singular at $\mathbf{x}=\mathbf{y}$. Therefore, this definition of quantum fields is also too singular as they are also not functions but rather operator valued distributions. This section will briefly discuss these distributions and how to derive well defined operators from them along with certain reasonable conditions placed on them. These operators will then be used to construct the Tomita-Takesaki modular operator framework.

\subsection{Operator Valued Distributions}

A Dirac delta function is almost always used by integrating it against a function. Similarly, 
an ordinary probability distribution is generally thought of as a mathematical object that needs to be integrated against some other function for it to be useful. Operator valued distribution are similar in that they may be thought of as a type of generalized function with values in linear operators. That is, an operator object $\Phi(x)$ is to be understood as being a distribution which when applied to a smooth test function $f(x)$ assigns the well-defined linear operator written as $\displaystyle\int_x f(x)\Phi(x)d^4x$. This is often colloquially called a \textit{smeared field} with properties to be presented below.

Consider an $n$-particle square integrable Hilbert space on four-dimensional space-time $x:=(x^0, \mathbf{x})$, $\mathcal{H} =
L_2(\mathbb{R}^{4n})$
with inner-product 
\begin{equation}
\begin{array}{c}
\braket{\psi | \phi}_n  
:= \\
\displaystyle\int_{\mathbb{R}^{4n}}
\psi^*(x_1,x_2,...,x_n)\phi(x_1,x_2,...,x_n)d^4x_1d^4x_2 \cdots d^4x_n,
\end{array}
\end{equation}
along with a generic \textit{one-particle-state} 
\begin{equation}
    f(x) \in L_2(\mathbb{R}^{4}).
\end{equation}
How does one combine such a state with an existing $n$-particle state $\psi_n(x_1,x_2,...,x_n)$.
One must start by creating a \textit{symmetric linear combination} (for Bosons) of all possible permutations of the space-time variables $x_i$ defined as
\begin{equation}
\begin{array}{c}
\rho_{n+1}(x_1, ...., x_{n+1}) = \dfrac{\sqrt{n+1}}{(n+1)!}\displaystyle\sum_{\bm\pi} f(x_1) \psi_n(x_2, ..., x_{n+1}), \\\\
where \;\; \bm\pi : \forall \; \text{permutations of} \; x_1, ..., x_{n+1} .
\end{array}
\end{equation}
These $n+1$-particle states may be created by operating with a \textit{creation operator} $\bm a^{\dagger}(f)$ of one particle states $f(x)$ on an $n$-particle state, i.e.
\begin{equation}
\begin{array}{c}
\bm a^{\dagger}(f) \psi_0 \longrightarrow \rho_1(x_1)=f(x_1)\psi_0, \\\\
\bm a^{\dagger}(f)\psi_1 \longrightarrow \rho_2(x_1,x_2)=\dfrac{f(x_1)\psi_1(x_2)+f(x_2)\psi_1(x_1)}{\sqrt{2}}, \\\\
etc.
\end{array}
\end{equation}
Single particle states may also be removed (effectively integrated out) from a multi-particle state using an \textit{annihilation operator} $\bm a(f)$ of one particle states. Using the notation
\begin{equation}
\begin{array}{c}
\chi_n(x_1, ..., x_n) \\\\=\sqrt{n+1} \displaystyle\int_{\mathbb{R}^4} f^*(x)\psi_{n+1}(x,x_1, ...., x_n)d^4x ,
\end{array}
\end{equation}
one has 
\begin{equation}
\begin{array}{c}
\bm a(f) \psi_1  \longrightarrow \chi_0 =\displaystyle\int_{\mathbb{R}^4} f^*(x)\psi_{1}(x)d^4x,  \\\\
\bm a(f) \psi_2  \longrightarrow \chi_1(x_1) =\sqrt{2}\displaystyle\int_{\mathbb{R}^4} f^*(x)\psi_{2}(x,x_1)d^4x, \\\\
etc.
\end{array}
\end{equation}

The Bosonic canonical commutation relations may now be calculated. Using the vacuum state $\psi_0$ defined by $\bm a(f) \psi_0=0 $, one has
\begin{equation}
\begin{array}{c}
\; [\bm a(f),\bm a^{\dagger}(g)]\psi_0 
=\bm a(f) \bm a^{\dagger}(g) \psi_0 - \bm a^{\dagger}(g)\bm a(f)\psi_0 \\\\=
\bm a(f) \bm a^{\dagger}(g) \psi_0
=\bm a(f) g(x)\psi_0\\\\ =  \displaystyle\int_{\mathbb{R}^4} f^*(x)g(x)d^4x =\braket{f | g} \psi_0.
\end{array}
\end{equation}
The general canonical commutation relations are
\begin{equation}
\begin{array}{c}
\; [\bm a^{\dagger}(f),\bm a^{\dagger}(g)]\psi = 0, \\\\ 
\; [\bm a(f),\bm a(g)]\psi = 0 ,\\\\
\; [\bm a(f),\bm a^{\dagger}(g)]\psi = \braket{f|g} \psi,
\end{array}
\end{equation}
which are sometimes simply written as
\begin{equation}
  \;[\bm a(f),\bm a^{\dagger}(g)]= \braket{f|g}, 
\end{equation}
along with the duality (adjoint) relation
\begin{equation}
\braket{\bm a(f) \psi | \phi} = \braket{\psi |\bm a^{\dagger}(f) \phi}    .
\end{equation}

Consider an orthonormal one particle basis $\{f_i\}$ often implicitly assumed in the standard physics literature where
\begin{equation}
\braket{f_i|f_j} = \int_{\mathbb{R}^4}  f^*_i(x) f_j(x) d^4x = \delta_{ij}   .
\end{equation}
The above commutation relation becomes
\begin{equation}
\; [\bm a(f_i),\bm a^{\dagger}(f_j)]\psi = \braket{f_i|f_j} \psi = \delta_{ij} \psi.
\end{equation}
By defining $a_i := \bm a(f_i)$ and $a^{\dagger}_j := \bm a^{\dagger}(f_j)$, one may write,
\begin{equation}
\; [a_i,a^{\dagger}_j]\psi = \delta_{ij} \psi,
\end{equation}
or as often seen,
\begin{equation}
\; [a_i,a^{\dagger}_j] = \delta_{ij}.
\end{equation}

The above formalism is consistent with the operator valued distributions $\bm a(x), \;\bm a^{\dagger}(x)$ defined as
\begin{equation}
\bm a^{\dagger}(g)  := \int_{\mathbb{R}^4} g(x) \bm a^{\dagger}(x)d^4x ,
\end{equation}
and
\begin{equation}
\bm a(f)  := \int_{\mathbb{R}^4} f^*(x) 
\bm a(x)d^4x ,
\end{equation}
if the operator valued distributions satisfy the commutation relation
\begin{equation}
 [\bm a(x), \bm a^{\dagger}(y)] = \delta(x-y) I ,  
\end{equation}
since
\begin{equation}
\begin{array}{c}
[\bm a(f), \bm a^{\dagger}(g)]\\\\=  \displaystyle\int_{\mathbb{R}^4}\int_{\mathbb{R}^4} f^*(x) g(y) [\bm a(x), \bm a^{\dagger}(y)]d^4x d^4y=\braket{f|g},
\end{array}
\end{equation}
as before.

The infinite degrees of freedom case of quantum field theory begins with the self-adjoint \textit{Segal field operator}
\begin{equation}
\Phi_s(f) = \dfrac{1}{\sqrt{2}}\biggl(a(f)+a^*(f)  \biggr)   , 
\end{equation}
and
\begin{equation}
\Phi_s(if) = -i\dfrac{1}{\sqrt{2}}\biggl(a(f)-a^*(f)  \biggr)    ,
\end{equation}
with
\begin{equation}
\begin{array}{c}
a^*(f)\Psi = \dfrac{1}{\sqrt{2}}\biggl( \Phi_s(f)-i\Phi_s(if)  \biggr)\Psi ,\\\\
a(f)\Psi = \dfrac{1}{\sqrt{2}}\biggl( \Phi_s(f)+i\Phi_s(if)  \biggr)\Psi.
\end{array}
\end{equation}
The field commutation relations are derived as
\begin{equation}
\begin{array}{c}
\;[\Phi_s(f), \Phi_s(g)]\Psi = 
\dfrac{1}{2}[a(f)+a^*(f), a(g)+a^*(g)] \Psi\\\\
=\dfrac{1}{2}[a(f), a^*(g)] \Psi +\dfrac{1}{2}[a^*(f), a(g)] \Psi \\\\
=\dfrac{1}{2}\braket{f,g} \Psi -\dfrac{1}{2}\braket{g,f} \Psi \\\\
= \dfrac{1}{2} \biggl(\braket{f,g}-\braket{f,g}^*  \biggr),
\end{array}
\end{equation}
and therefore
\begin{equation}
  \;[\Phi_s(f), \Phi_s(g)] =   i \text{Im}\braket{f,g}.
\end{equation}
where $\braket{f,g}$ is often a covariant scalar product involving Greens function style propagators.

Since $\Phi_s(f)$ is self-adjoint, the following Weyl operators are well-defined and unitary 
\begin{equation}
  W(f)= e^{i\Phi_s(f)}. 
\end{equation}
The Weyl commutation relations are
\begin{equation}
\begin{array}{c}
  W(f)W(g) =  e^{-i\text{Im}\braket{f,g}/2} W(f+g),\\\\
  W(f)W(g)= e^{-i\text{Im}\braket{f,g}}W(g)W(f) ,
\end{array}
\end{equation}
and will be used in the analysis of the Unruh-DeWitt detector model below.

We know define a $\bm{C}^*$-algebra of quantum fields on Minkowski spacetime $\mathcal{M}$, denoted as $\mathcal{A}(\mathcal{M})$. The quantum field generators are denoted as $\Phi$. Let $f \in C_0^{\infty}(\mathcal{M})$ be smooth complex functions with compact support in $\mathcal{M}$. The elements of $\mathcal{A}(\mathcal{M})$ are formed from a \textit{sum of operator products} \begin{equation}\Phi(f_1)\Phi(f_2)\Phi(f_3)\cdots,
\end{equation}
of \textit{smeared fields} 
\begin{equation}
\Phi(f) = \int_{\mathcal{M}} \phi(x)f(x)d\mu(x),   
\end{equation}
where $d\mu(x)$ is an appropriately chosen differential volume element in $\mathcal{M}$.

These  quantum fields have the following properties:

\noindent \textit{Linearity}:
\begin{equation}
 \Phi(\alpha f_1 +\beta f_2) =\alpha  \Phi(f_1) +\beta  \Phi(f_2), \;\;\;   \alpha, \beta \in \mathbb{C}  
\end{equation}

\noindent \textit{Hermiticity}:
\begin{equation}
 \Phi(f)^* =  \Phi(\bar{f})
\end{equation}

\noindent \textit{Satisfy field equations such as the Klein-Gordon equation}:
\begin{equation}
 \Phi\bigl((\Box+m^2)f \bigr) =0
\end{equation}

\noindent \textit{Obey covariant commutation relations}:
\begin{equation}
\; [\Phi(f),\Phi(g)] =i \braket{f,g}
\end{equation}

Finally, consider a causally convex bounded region $\mathcal{O} \subset \mathcal{M}$ and the local sub-algebra with support in this causal domain, $\mathcal{A}(\mathcal{O}) \subset \mathcal{A}(\mathcal{M})$. Causality here is  defined as follows.
If $\mathcal{O}_1$ and $\mathcal{O}_2$
are causally disjoint regions,  one must have
\begin{equation}
\;[\mathcal{A}(\mathcal{O}_1) , \mathcal{A}(\mathcal{O}_2)]=\{0\},
\end{equation}
or explicitly
\begin{equation}
\begin{array}{c}
\forall A_1 \in \mathcal{A}(\mathcal{O}_1) \;\&\;\forall A_2 \in \mathcal{A}(\mathcal{O}_2) \; \longrightarrow
\;[A_1, A_2]=0.
\end{array}
\end{equation}
Further details may be found in \cite{Bratteli1997, Haag1992}.

\subsection{Tomita-Takesaki Modular Operators}

We begin by discussing dual systems where in this context are those systems that have mutually commuting von Neumann algebras of operator observables. 
Consider a set bounded linear operators on a Hilbert space $\mathcal{H}$, $\mathcal{B(H)}$. The algebra of one system $\mathcal{A} \subset \mathcal{B(H)}$ is a $\bm{C}^*$-algebra with a commutant $\mathcal{A}'$, the algebra of the other system,  being the set of elements in $\mathcal{A}' \subset \mathcal{B(H)}$ commuting with $\mathcal{A}$. If the double commutant returns one to $\mathcal{A}$, i.e. $\mathcal{A}''=\mathcal{A}$, both $\mathcal{A}$ and its dual $\mathcal{A}'$ are von Neumann algebras. Von Neumann showed that this bicommutant definition is equivalent to the algebra $\mathcal{A}$ being closed with respect to the weak topology on $\mathcal{B(H)}$. 

The duality between these two von Neumann algebras may be explicitly formulated using Tomita-Takesaki modular operators (with details found in the Appendix). Let $|\Omega \rangle  \in \mathcal{H} $ be a cyclic and separating vector for $\mathcal{A}$ meaning that the elements in  $\mathcal{A}$ can create a space dense in $\mathcal{H}$ by acting on $|\Omega \rangle  \in \mathcal{H}$. The Reeh-Schlieder theorem states that the vacuum state of some quantum field theory is cyclic meaning that $\mathcal{A}(\mathcal{O})\ket{\Omega}$ is dense in the whole Hilbert space $\mathcal{H}$ for any open region $\mathcal{O}$ in Minkowski space. Further, $\mathcal{A}(\mathcal{O}^*)\ket{\Omega}$ is also dense in $\mathcal{H}$, where $\mathcal{O}^*$ is the causal compliment to $\mathcal{O}$.   
Tomita-Takesaki theory states that there exists an anti-linear map $ S: \mathcal{H} \rightarrow \mathcal{H},$ such that $ S A |\Omega \rangle  = A^* |\Omega \rangle, \forall A \in \mathcal{A} $. $S$ has a polar decomposition given by $S = J \Delta^{1/2} =  \Delta^{-1/2} J, \,\, \Delta = S^{*} S$ where the modular conjugation operator $J$ has the following properties,
\begin{equation}
    \begin{array}{cc}
        J \Delta ^{\frac{1}{2}} J = \Delta ^{-\frac{1}{2}}, \;\;\;
        J^2 =I,  \\\\
        J |\Omega \rangle  = |\Omega \rangle,  \;\;\;
        J \mathcal{A} J = \mathcal{A}' .
    \end{array}
\end{equation}
 Note that the modular conjugation operator $J$ directly provides the duality between the algebras $\mathcal{A}$ and $\mathcal{A}'$.

The above framework was explicitly formulated for a supersymmetric quantum mechanical system in \cite{Chatterjee1}. A brief review and extension of those results will be given in Sec. IV. below. For the scalar field theory with Unruh-DeWitt qubit detectors, we will need some further elaboration following \cite{summers1987I, summers1987II, Fabritiis2023}.  Let $f \in C_0^{\infty}(\mathcal{O})$ be smooth complex functions with compact support in an open region of Minkowski spacetime $\mathcal{O}$. Denote the space of such test functions as $S(\mathcal{O})$. Consider now the symplectic complement of $S(\mathcal{O})$, denoted by 
$S(\mathcal{O}')$, as the space of test functions with vanishing functional inner product with those in  $S(\mathcal{O})$, 
\begin{equation}
\begin{array}{c}
 \braket{f,g} = \displaystyle\int f(x) G(x-y) g(y) d^4x d^4y =0,\\\\
 f \in S(\mathcal{O}), \; g \in S(\mathcal{O}').
 \end{array}
\end{equation}
where $G(x-y)$ is a propagator such as a Pauli-Jordon Green's function.
This allows one to restate the causality condition for operators stated above as
\begin{equation}
 \; [\phi(f),\phi(g)]=0, \;\;\;
 \forall f \in S(\mathcal{O}) \;\text{and}\; \forall \; g \in S(\mathcal{O}').
\end{equation}

Now consider a von Neumann algebra $\mathcal{A}(S(\mathcal{O}))$ of Weyl style operators $A=e^{i\phi(f)}$ with $f \in S(\mathcal{O})$. A properly defined modular conjugation operator $J$ should take this algebra to its commutant algebra $J \mathcal{A} J = \mathcal{A}'$. Using Haag duality \cite{Haag1992} for free Bose fields, one can state that $\mathcal{A}'(S(\mathcal{O})) = \mathcal{A}(S(\mathcal{O}'))$ such that
the commutant $\mathcal{A}'$ is seen to be equivalent to the algebra $\mathcal{A}$ defined on the symplectic complement space of test functions. Using this fact, the authors of \cite{Fabritiis2023} define the action of the modular conjugation operator $J$ as $Je^{i\phi(f)}J=e^{-i\phi(jf)}$ where $f \in S(\mathcal{O})$ and defining $jf \in S(\mathcal{O}')$. They have effectively lifted the Tomita-Takesaki modular conjugation operator $J$ to the space of test functions  $S(\mathcal{O})$ with the definition of $j$. Finally, for 
$f \in S(\mathcal{O}), \; g \in S(\mathcal{O}')$, they derive the following vacuum state expectation value using the Weyl commutation relations stated above,
\begin{equation}
 \bra{0}e^{i\phi(f)}e^{i\phi(g)}\ket{0} = \bra{0}e^{i\phi(f+g)}\ket{0} = e^{-\frac{1}{2}||f+g||^2} ,\label{WeylExpectation}
\end{equation}
where $||h||=\braket{h,h}$.
This result will be used in the analysis of the Unruh-DeWitt qubit detector below.

\section{Concurrence and the Bell-CHSH Inequality}

Quantum concurrence is a measure used in quantum information theory to quantify the amount of entanglement of a bipartite qubit state. For a pure bipartite state $\ket{\psi}$, the concurrence $C(\ket{\psi})$ is given by the overlap \cite{Woot2001, Beng2017}
\begin{equation}
  C = |\langle \tilde{\psi}|\psi\rangle|  ,  
\end{equation}
where the "spin-flipped" state $|\tilde{\psi}\rangle$
is defined by
\begin{equation}
|\tilde{\psi}\rangle = (\sigma^y \otimes \sigma^y)|\psi^*\rangle  ,
\end{equation}
 where is the complex conjugate of $|\psi\rangle$.

For a mixed state described by a density matrix $\rho$, the calculation is somewhat more complex.  First, define the spin-flipped density operator as
\begin{equation}
\tilde{\rho} =  (\sigma^y \otimes \sigma^y) \rho^* (\sigma^y \otimes \sigma^y)  ,
\end{equation}
where the complex conjugation is taken in the computational basis. Now, create the following Hermitian matrix
\begin{equation}
R =  \sqrt{\rho}\tilde{\rho}\sqrt{\rho} .
\end{equation}
 Next,  find the eigenvalues $ \lambda_i$ of the matrix $R$. The concurrence $C(\rho)$ may now be calculated as 
\begin{equation}
   C(\rho) = \max(0, \sqrt{\lambda_1} - \sqrt{\lambda_2} - \sqrt{\lambda_3} - \sqrt{\lambda_4}),
\end{equation}
where \(\lambda_1, \lambda_2, \lambda_3,\) and \(\lambda_4\) are the eigenvalues of \(R\) in decreasing order. Concurrence ranges from 0 to 1, where 0 indicates that the state is not entangled (a separable state), and 1 indicates maximal entanglement. Note that this definition is sometimes stated using the non-Hermitian matrix $R'=\rho \tilde{\rho}$. 

The Bell-CHSH inequality involves measurements on two qubits and is given by:
\begin{equation}
S = \left| E(a, b) + E(a, b') + E(a', b) - E(a', b') \right| \leq 2,
\end{equation}
where \( E(a, b) \) is the expectation value of the product of the outcomes when measurements are made along directions \( a \) and \( b \). The parameters \( a \) and \( a' \) are two different measurement settings for one qubit, and \( b \) and \( b' \) are two different measurement settings for the other qubit. In quantum field theory, \( E(a, b) \) represents the expectation value of the product of the measurement outcomes for field operators at measurement settings \( a \) and \( b \) respectively, 
\begin{equation}
E(a, b) = \langle \psi | \hat{A}(a) \hat{B}(b) | \psi \rangle.
\end{equation}
Here, \( \hat{A}(a) \) and \( \hat{B}(b) \) are the measurement operators corresponding to the measurement settings \( a \) and \( b \) for two space-like separated regions (disjoint causal sets) and \( |\psi\rangle \) is the entangled state of the field. The Bell-CHSH inequality can also be expressed using the Bell-CHSH field operator \(\hat{\mathcal{B}}\) defined as
\begin{equation}
\hat{\mathcal{B}} = \hat{A}(a) \hat{B}(b) + \hat{A}(a) \hat{B}(b') + \hat{A}(a') \hat{B}(b) - \hat{A}(a') \hat{B}(b').
\end{equation}
The Bell-CHSH inequality states that for any local hidden variable theory, the expectation value \( \langle \mathcal{B} \rangle \) satisfies the inequality
\begin{equation}
|\langle \mathcal{B} \rangle| := \left| \langle \psi | \hat{\mathcal{B}} | \psi \rangle \right|\leq 2,
\end{equation}
where \(|\psi\rangle\) is the entangled state of the system.

Quantum systems have shown that for certain entangled states, \textit{this bound can be violated} and the maximum value it can reach is \( 2\sqrt{2} \).
For a given concurrence \( C \), the maximum violation of the Bell-CHSH inequality is given by
\begin{equation}
|\langle \mathcal{B} \rangle|_{\text{max}} = 2\sqrt{1 + C^2}.
\end{equation}
For the quantum concurrence of $C = 1$ for a maximally entangled state, one has \( \langle \mathcal{B} \rangle_{\text{max}} = 2\sqrt{2} \)  indicating the maximum quantum violation known as Tsirelson's bound.

\section{A Supersymmetric Quantum Mechanical System}

We briefly review and extend some previous work of the author \cite{Chatterjee1} with further details and references therein. A similar procedure will be used for the Unruh-DeWitt qubit detector model discussed below.

Recall that the Hamiltonian of an electron with motion restricted to a two-dimensional plane in a transverse magnetic field is given by
\begin{equation}
H = \dfrac{1}{2m} \mathbf{\Pi} ^2 -\boldsymbol{\mu} \cdot\mathbf{B},
\end{equation}
where $ \mathbf{\Pi} = \mathbf{p} - \dfrac{e}{c} \mathbf{A}$ is the Peierls momenta.
This Hamiltonian may be written as 
\begin{equation}
H^a = \hbar \omega \left( a^\dagger a  + \dfrac{1}{2} -\dfrac{\hat{\sigma}_z}{2}  \right),
\end{equation}
using the standard quantization prescription initiated by introducing creation and annihilation operators related to the quantized Peierls momenta with frequency $\omega = eB/mc$,
\begin{equation}
\begin{array}{c}
[\Pi_x , \Pi_y] = i \dfrac{e\hbar}{c} B ,
\\\\
a = \sqrt{\dfrac{c}{2e\hbar B}} (i\Pi_x - \Pi_y ), 
\\\\
a^\dagger = \sqrt{\dfrac{c}{2e\hbar B}} (-i\Pi_x - \Pi_y ) .
\end{array}
\end{equation}
The Hilbert space is $\mathcal{H} = \mathcal{F} \otimes \mathbb{C}^2$ where $ |n \rangle \in \mathcal{F}$ is the Bosonic Fock space and $\alpha 
\begin{pmatrix}
1 \\
0
\end{pmatrix} 
+\gamma \begin{pmatrix}
0 \\
1
\end{pmatrix} 
 \in \mathbb{C}^2$. 
The degeneracy of the associated Landau energy levels along with the unique ground state energy leads one to a SUSY quantum mechanical model \cite {Chatterjee1, Ezawa2008} where the supersymmetric charges are given by
\begin{equation}
Q^a = a  \otimes \sigma_- = 
\begin{pmatrix}
0 & 0 \\
a & 0
\end{pmatrix}, \;\;
\;\;
Q^{a \dagger} = a^\dagger  \otimes \sigma_+ =
\begin{pmatrix}
0 & a^\dagger \\
0 & 0
\end{pmatrix},
\end{equation}
with Hamiltonian
\begin{equation}
H^a  = \hbar \omega\{ Q^a, Q^{a \dagger} \} \label{Hamil}
\end{equation}
and corresponding supermultiplet states
\begin{equation}
\left\{ \Phi^{k}_{\uparrow} ,\Phi^{k-1}_{\downarrow} \right\} =
\left\{
\begin{pmatrix}
|k \rangle \\
0
\end{pmatrix} ,
\begin{pmatrix}
0  \\
|k-1 \rangle
\end{pmatrix}
\right\},
\end{equation}
with energy $E^a_k  = \hbar \omega k$. 

The well known degeneracy of each Landau level (related to the angular momentum of cyclotron motion) and the degeneracy "filling factor" for finite systems suggests another structure creating this degeneracy. It is shown in \cite{Chatterjee1} that such a system may be interpreted as one of the two cases of positive and negative $z$-directions of the transverse magnetic field. These two systems will be identified by the superscripts '$a$' and '$b$'. The '$b$' operators commute with '$a$' operators such that supersymmetric algebraic structure of the  '$b$' system is similar to the '$a$' system, i.e.
\begin{equation}
Q^{b \dagger} = b^\dagger  \otimes \sigma_- =
\begin{pmatrix}
0 & 0 \\
b^\dagger & 0
\end{pmatrix}, \;\;
\\
Q^{b} = b  \otimes \sigma_+ = 
\begin{pmatrix}
0 & b \\
0 & 0
\end{pmatrix},
\end{equation}
with
\begin{equation}
H^b  =\hbar \omega \{ Q^b , Q^{b \dagger} \} 
\end{equation}
and the two states 
\begin{equation}
\left\{ \Phi^{l-1}_{\uparrow} ,\Phi^{l}_{\downarrow} \right\} =
\left\{
\begin{pmatrix}
|l-1 \rangle \\
0
\end{pmatrix} ,
\begin{pmatrix}
0  \\
|l \rangle
\end{pmatrix}
\right\},
\end{equation}
forming the supermultiplet with energy $E^b_l =  \hbar \omega l $. 
 As these operators commute with the Hamiltonian (\ref{Hamil}), the explicit degeneracy of the now composite Bosonic Fock space $\mathcal{F}$ of $\mathcal{H} = \mathcal{F} \otimes \mathbb{C}^2$, $|n,m \rangle \in \mathcal{F}$ can be written as  
\begin{equation}
|n,m \rangle = \dfrac{{a^\dagger}^n {b^\dagger}^m}{\sqrt{n!m!}} |0,0 \rangle , n,m \in \mathbb{Z}^{0+} .
\end{equation}

The duality between these two systems of commuting algebras may be generated by Tomita-Takesaki modular operators.
The key feature for our purposes is that the modular conjugation operator $J$ takes an algebra $\mathcal{A}$ into its commutant $\mathcal{A}'$. The other operators of the Tomita-Takesaki framework such as $S$ and $\Delta$ are fully discussed in \cite{Chatterjee1}. 
 The von Neumann algebras considered here will have elements of the form of unitary Weyl operators of the supercharge generators $Q^a, Q^{a \dagger}, Q^b, Q^{b \dagger}$,
\begin{equation}
\begin{array}{c}
\Big\{ \exp[i(\alpha Q^a +\beta Q^{a \dagger} )] \Big\} \subset \mathcal{A} ,\\\\
\Big\{ \exp[i(\gamma Q^b +\delta Q^{b \dagger} )] \Big\} \subset \mathcal{A}' .
\end{array}
\end{equation}
The $*$-operation of the $C^*$-algebra structure is given by the adjoint operation $\dagger$. The modular conjugation operator $J: \mathcal{F} \otimes \mathbb{C}^2 \rightarrow \mathcal{F} \otimes \mathbb{C}^2 $ is defined by
\begin{equation}
J \left[ |n , m \rangle \otimes 
\begin{pmatrix}
\alpha\\
\beta 
\end{pmatrix}  \right] \\\\
= 
\left[ |m, n \rangle \otimes 
\begin{pmatrix}
\bar{\beta} \\
\bar{\alpha}
\end{pmatrix} \right].
 \label{SUSY-J}
\end{equation}
For example, to show that $J$ maps $\mathcal{A}$ into $\mathcal{A}'$, i.e. $J \mathcal{A} J = \mathcal{A}'$,
note that
\begin{equation}
\begin{array}{c}
J Q^a J 
\left[ |n , m \rangle \otimes 
\begin{pmatrix}
\alpha\\
\beta 
\end{pmatrix}  \right]
=
J [a  \otimes \sigma_{-} ]
\left[ |m, n \rangle \otimes 
\begin{pmatrix}
\bar{\beta} \\
\bar{\alpha}
\end{pmatrix} \right] \\\\ = 
J \left[ \sqrt{m} \ket{m-1,n} \otimes \begin{pmatrix}
0 \\
\bar{\beta}
\end{pmatrix}
\right]
  \\\\=
  \sqrt{m} \ket{n,m-1} \otimes \begin{pmatrix}
\beta \\
0
\end{pmatrix} = [b \otimes \sigma_{+} ]\left[ |n , m \rangle \otimes 
\begin{pmatrix}
\alpha\\
\beta 
\end{pmatrix}  \right]
\\\\=
Q^b \left[ |n , m \rangle \otimes 
\begin{pmatrix}
\alpha\\
\beta 
\end{pmatrix}  \right] .
\end{array}
\end{equation}
A similar calculation holds for the other operators.

Denoting the composite Bosonic part as $\mathcal{F}=\mathcal{F}^a \otimes \mathcal{F}^b$, consider the following entangled supermultiplet state in $ \mathcal{F} \otimes \mathbb{C}^2$, 
\begin{equation}
| \Phi \rangle = \alpha \, | k ,\, l-1 \rangle \otimes \begin{pmatrix}
1 \\
0
\end{pmatrix} + \beta \, | l-1 ,\, k \rangle \otimes \begin{pmatrix}
0 \\
1
\end{pmatrix} .
\end{equation}
The concurrence of $C(| \Phi \rangle )$ is given by the absolute value of the expectation value of the modular conjugation operator in this state, 
\begin{equation}
C(| \Phi\rangle ) = | \langle \Phi | J | \Phi \rangle |, \label{conc}
\end{equation}
as first introduced in \cite{Chatterjee1}.
Using (\ref{SUSY-J}), we have
\begin{equation}
 J | \Phi \ \rangle =  \bar{\alpha} | l-1 ,\,k \rangle \otimes \begin{pmatrix}
0 \\
1
\end{pmatrix} + \bar{\beta} |k ,\, l-1 \rangle \otimes \begin{pmatrix}
1 \\
0
\end{pmatrix} ,
\end{equation} 
and therefore, 
\begin{equation}
| \langle \Phi | J | \Phi \rangle |= 2 |\alpha \beta |,
\end{equation}
which is the concurrence using traditional methods. For a maximally entangled state with $\alpha = \beta = 1/\sqrt{2}$, $C(| \Phi \rangle ) = 1$ as required. The concurrence relation (\ref{conc}) has given a physical meaning to the modular conjugation operator as a quantitative measure of entanglement for a bi-partite supermultiplet state. As described in \cite{Uhlmann2000}, an anti-linear, anti-unitary operator is the key driving force behind the concurrence calculation of bipartite systems and it is now understood in the Tomita-Takesaki framework as the modular conjugation operator $J$. 

\section{Unruh-DeWitt Qubit Detectors}

\subsection{Interaction with a Klein-Gordon Field}

The Unruh-DeWitt particle detector model \cite{Unruh1976, DeWitt1979, Tjoa2023} is a simple quantum mechanical system where a qubit, such as a two-level atom, interacts with a scalar quantum field. The detector coupling to the quantum field allows it to absorb or emit the quanta of the scalar field leading to transitions between its energy levels. The Unruh-DeWitt detector was first used to study the Unruh effect where an observer accelerating through empty space (a vacuum) will detect a thermal bath of particles even though an inertial observer would see none.  The model has been used to study quantum fields in curved spacetime, such as those near a black hole, and phenomena like Hawking radiation where black holes emit thermal radiation due to quantum effects.
The context here is more related to the framework of a quantum channel consisting of two localized qubit systems that communicate through the scalar quantum field. 

The Hilbert spaces of the detector and the field are denoted as \( \mathcal{H}_D \) and \( \mathcal{H}_F \), respectively. The total Hilbert space is \( \mathcal{H} = \mathcal{H}_D \otimes \mathcal{H}_F \).
The detector Hamiltonian \( H_D \) is:
\begin{equation}
H_D = \frac{1}{2}  \omega (\sigma^z+I_D) \otimes I_F,
\end{equation}
where \( \sigma^z \) is the Pauli matrix acting on \( \mathcal{H}_D \), and \( I_F \) is the identity operator on \( \mathcal{H}_F \).
The field Hamiltonian \( H_F \) is:
\begin{equation}
H_F = I_D \otimes \int d^3x \left( \frac{1}{2} \pi^2 + \frac{1}{2} (\nabla \phi)^2 + \frac{1}{2} m^2 \phi^2 \right),
\end{equation}
where \( I_D \) is the identity operator on \( \mathcal{H}_D \).
The traditional interaction Hamiltonian \( H_I \) is:
\begin{equation}
\begin{array}{c}
H_I =  \left[\sigma_{+} e^{i\omega \tau} + \sigma_{-} e^{-i\omega \tau}\right] \otimes \phi(x(\tau)),
\end{array}
\end{equation}
where \( \sigma_+ \) and \( \sigma_- \) are the raising and lowering operators acting on \( \mathcal{H}_D \), and \( \phi(x(\tau)) \) is the field operator evaluated along the detector's trajectory ($\tau$ is the proper time) acting on \( \mathcal{H}_F \).
The term
\begin{equation}
\mu(\tau) = \sigma_{+} e^{i\omega \tau} + \sigma_{-} e^{-i\omega \tau} ,
\end{equation}
is sometime referred to as the monopole moment of the detector such that we may simply write the interaction term as $\mu(\tau) \otimes \phi(x)$ 
or in terms of smeared fields as
\begin{equation}
H_I = \int dV f(x) \biggl(\mu(\tau) \otimes \phi(x)\biggr),
\end{equation}
where $f(x) \in C_0^{\infty}(\mathcal{M})$ is a spacetime smearing test function and $dV$ is an invariant differential volume element of $\mathcal{M}$. Unitary evolution in the interaction picture is given by the time-ordered expression
\begin{equation}
 U=\mathcal{T}_{\tau}\exp\left[-i\int dV f(x) \biggl(\mu(\tau) \otimes \phi(x)\biggr)\right] . \end{equation}

 The time evolution may be explicitly evaluated in a non-perturbative manner, following \cite{Tjoa2023}, by simplifying the dynamics of the model. One such simplification is to consider a gapless detector where one sets $\omega =0$ such that $H_D=0$. The monopole moment term $\mu(\tau)$ becomes a constant operator allowing the unitary evolution to be written explicitly using a smeared scalar field,
\begin{equation}
\begin{array}{c}
 U=\mathcal{T}_{\tau}\exp\left[- \biggl(\mu \otimes i\displaystyle\int dV f(x)\phi(x)\biggr)\right]  \\\\= 
\mathcal{T}_{\tau}\exp\left[- i\biggl(\mu\otimes \phi(f) \biggr)\right] .
  \end{array}
 \end{equation}
After a Magnus expansion, one is left with the simplified expression
\begin{equation}
U=e^{-i\Delta}e^{-i\mu \otimes \phi(f)} ,
\end{equation}
where $e^{-i\Delta}$ is a global phase factor given explicitly in \cite{Tjoa2023} that does not affect the detector dynamics.

 Another simplification to the dynamics that allows for an explicit evaluation of the evolution operator is to choose an interaction Hamiltonian that commutes with $H_D$. Choosing the monopole operator to be the constant Pauli matrix operator found in $H_D$, $\mu=\sigma^z$, one has 
\begin{equation}
H_I = \int dV f(x) \biggl(\sigma^z \otimes \phi(x)\biggr),
\end{equation}
or after a Magnus expansion,
\begin{equation}
U=e^{-i\Delta}e^{-i\sigma^z \otimes \phi(f)} ,
\end{equation}
where $e^{-i\Delta}$ is the same global phase factor as in the previous case. This has been called the \textit{pure dephasing model} in \cite{Tjoa2023}.

\subsection{Quantum Concurrence for Entangled Unruh-DeWitt Qubits}

In \cite{Guedes2024}, the authors calculated the Bell-CHSH inequality for an entangled state of two Unruh-DeWitt qubit detectors interacting with a Bosonic quantum field using the Tomita-Takesaki modular theory of von Neumann algebras of Weyl operators for the scalar field. They used a formalism explained in Sec. II above where causality is restated by introducing a symplectic complement of the original set of test functions (for the operator valued distributions). Because the Bosonic Pauli-Jordon distribution function disappears between any two such test functions $f \in S(\mathcal{O}), \; g \in S(\mathcal{O}')$, one has $[\phi(f), \phi(g)]=0$ and the vacuum expectation of Weyl operators may be evaluated to be $\bra{0}e^{i\phi(f)}e^{i\phi(g)}\ket{0} = \bra{0}e^{i\phi(f+g)}\ket{0} = e^{-\frac{1}{2}||f+g||^2} $.

Here, we extend the modular operator approach to the Unruh-DeWitt detector Hilbert space in a manner similar to the approach of the supersymmetric model in Sec. IV. Thereafter, we calculate the quantum concurrence using the Tomita-Takesaki modular conjugation operator and show it is consistent with the results of  \cite{Guedes2024} in the sense of its relation to the maximum Bell-CHSH inequalities as described in Sec. III. 

Let $\ket{0}$ denote the vacuum of the Klein Gordon scalar field. The ground state and excited state of the Unruh-DeWitt detector are simply the qubit states $\begin{pmatrix}
0 \\
1
\end{pmatrix} $ and $\begin{pmatrix}
1 \\
0
\end{pmatrix} $ with energies $0$ and $\omega$. Following the tradition of quantum information analysis, we begin with an entangled pair of localized qubits from two separated systems $A$ and $B$, often referred to as Alice and Bob.  Let Alice and Bob each locally couple their two-level Unruh-DeWitt entangled detectors to a scalar quantum field $\phi$. Consider the initial entangled state 
\begin{equation}
\begin{array}{c}
\ket{\psi}_0=
    \dfrac{1}{\sqrt{1+r^2}}
 \left[\begin{pmatrix}
0 \\
1
\end{pmatrix} \otimes \begin{pmatrix}
0 \\
1
\end{pmatrix} + r\begin{pmatrix}
1 \\
0
\end{pmatrix} \otimes \begin{pmatrix}
1 \\
0
\end{pmatrix}  \right] \otimes \ket{0},
\end{array}
\end{equation}
where $r$ is a parameter to control the entanglement. Note that $r=1$ gives the maximally entangled state whereas $r=0$ reduces to a separable state. 
We now specialize to a pure dephasing model \cite{Tjoa2023} as the one described above with a unitary evolution operator of (and dropping the global phase factor)
\begin{equation}
    U= e^{-i\sigma^z_A \otimes \phi(f_A) } \otimes e^{-i\sigma^z_B \otimes \phi(f_B) }.
\end{equation}
The evolved wave-function becomes \cite{Guedes2024}
\begin{equation}
\begin{array}{c}
\ket{\psi}=
    \dfrac{1}{\sqrt{1+r^2}} 
\left[\begin{pmatrix}
0 \\
1
\end{pmatrix} \otimes \begin{pmatrix}
0 \\
1
\end{pmatrix} \otimes e^{i\phi(f_A+f_B)}\ket{0} 
\right. \\\\
\left.+ r\begin{pmatrix}
1 \\
0
\end{pmatrix} \otimes \begin{pmatrix}
1 \\
0
\end{pmatrix}  \otimes e^{-i\phi(f_A+f_B)}\ket{0} \right].
\end{array}
\end{equation}

The von Neumann algebras considered here will have elements of the form of unitary Weyl operators composed of Pauli matrices obeying the Pauli matrix algebra of $ \sigma^i\sigma^j=\delta^{ij}+ i\epsilon^{ijk}\sigma^k,  \;i,j=x,y,z$,
\begin{equation}
\begin{array}{c}
\Big\{ \exp[i(\sigma^i_A \otimes I_B)] \Big\} \subset \mathcal{A},\;\;\;(Alice),\\\\
\Big\{ \exp[i(I_A \otimes \sigma^i_B )] \Big\} \subset \mathcal{A}', \;\;\;(Bob).
\end{array}
\end{equation}
These act on the Hilbert space $\mathcal{H}=\mathbb{C}^2 \otimes \mathbb{C}^2$.
The $*$-operation of the $C^*$-algebra structure is given by the adjoint operation $\dagger$. 
The modular conjugation operator here is analogous to the supersymmetric case of Sec IV. Let $J_{AB}: \mathbb{C}_A^2 \otimes \mathbb{C}_B^2 \rightarrow \mathbb{C}_A^2 \otimes \mathbb{C}_B^2  $ where explicitly 
\begin{equation}
J_{AB}\left[
\begin{pmatrix}
\alpha \\
\beta
\end{pmatrix} \otimes
\begin{pmatrix}
\gamma \\
\delta
\end{pmatrix}  \right]
=
\begin{pmatrix}
\bar{\delta} \\
\bar{\gamma}
\end{pmatrix} \otimes
\begin{pmatrix}
\bar{\beta} \\
\bar{\alpha}
\end{pmatrix} .
\end{equation}
The symmetry operation mapping the observables of system $A$ to $B$ can be seen, for example, by
\begin{equation}
\begin{array}{c}
J_{AB}[\sigma_y \otimes I] J_{AB} \left[\begin{pmatrix}
\alpha \\
\beta
\end{pmatrix} \otimes
\begin{pmatrix}
\gamma \\
\delta
\end{pmatrix} \right] \\\\
= J_{AB}[\sigma_y \otimes I] \left[\begin{pmatrix}
\bar{\delta} \\
\bar{\gamma}
\end{pmatrix} \otimes
\begin{pmatrix}
\bar{\beta} \\
\bar{\alpha}
\end{pmatrix}  \right]\\\\
=J_{AB} \left[\begin{pmatrix}
-i\bar{\gamma} \\
i\bar{\delta}
\end{pmatrix} \otimes
\begin{pmatrix}
\bar{\beta} \\
\bar{\alpha}
\end{pmatrix}\right]
=
\begin{pmatrix}
\alpha \\
\beta
\end{pmatrix} \otimes
\begin{pmatrix}
-i\delta \\
i\gamma
\end{pmatrix}\\\\
=[I \otimes \sigma_y]
\left[
\begin{pmatrix}
\alpha \\
\beta
\end{pmatrix} \otimes
\begin{pmatrix}
\gamma \\
\delta
\end{pmatrix}
\right].
\end{array} 
\end{equation}

As the concurrence of $C\bigl(\ket{\psi}\bigr)$ is given by the absolute value
of the expectation value of the modular conjugation operator in the state $\ket{\psi}$ following \cite{Chatterjee1}, we begin by
operating with $J_{AB} \otimes I$ on $\ket{\psi}$,
\begin{equation}
 \begin{array}{c}
(J_{AB} \otimes I)\ket{\psi} = \dfrac{1}{\sqrt{1+r^2}} \\\\
\times \left[\begin{pmatrix}
1 \\
0
\end{pmatrix} \otimes \begin{pmatrix}
1 \\
0
\end{pmatrix} \otimes e^{i\phi(f_A+f_B)}\ket{0} \right.\\\\
\left.
+ r\begin{pmatrix}
0 \\
1
\end{pmatrix} \otimes \begin{pmatrix}
0 \\
1
\end{pmatrix}  \otimes e^{-i\phi(f_A+f_B)}\ket{0}
\right].
 \end{array}
\end{equation}
Finally, the quantum concurrence is calculated to be
\begin{equation}
\begin{array}{c}
    C\bigl(\ket{\psi}\bigr) = |\bra{\psi}(J_{AB} \otimes I)\ket{\psi}| \\\\= 
    \dfrac{1}{1+r^2} \biggl[r\braket{0|e^{-2i\phi(f_A+f_B)}|0}+r\braket{0|e^{2i\phi(f_A+f_B)}|0} \biggr] \\\\=
    \dfrac{2r}{1+r^2}e^{-2||f_A+f_B||^2},
\end{array} \label{mainresult}
\end{equation}
using (\ref{WeylExpectation}).
Clearly, the scalar quantum field $\phi$ has a damping effect on the entanglement of the system as the zero field ($\phi \rightarrow 0$) isolated entangled Unruh-DeWitt detector state 
\begin{equation}
    \dfrac{1}{\sqrt{1+r^2}}
 \left[\begin{pmatrix}
0 \\
1
\end{pmatrix} \otimes \begin{pmatrix}
0 \\
1
\end{pmatrix} + r\begin{pmatrix}
1 \\
0
\end{pmatrix} \otimes \begin{pmatrix}
1 \\
0
\end{pmatrix}  \right] ,
\end{equation}
has a concurrence given by
\begin{equation}
C_0=\dfrac{2r}{1+r^2}.
\end{equation}

Our concurrence expression (\ref{mainresult})
is consistent with the Bell-CHSH inequality result found in \cite{Guedes2024} ($\alpha,\beta, \alpha',\beta'$ are the four Bell angles)
\begin{equation}
\begin{array}{c}
\braket{\mathcal{B}}=
    \dfrac{2r}{1+r^2}e^{-2||f_A+f_B||^2} \times \\\\
\;[\cos(\alpha+\beta)+\cos(\alpha'+\beta)+\cos(\alpha+\beta')-\cos(\alpha'+\beta')],
\end{array}
\end{equation} 
when used in junction with the maximum violation relation of the Bell-CHSH inequality for a specific concurrence given in Sec. III. The Bell angles can be chosen such that the angular part above can be  maximized to $2\sqrt{2}$ such that 
\begin{equation}
\braket{\mathcal{B}}=2\sqrt{2} \left(
    \dfrac{2r}{1+r^2}e^{-2||f_A+f_B||^2} \right),
\end{equation} 
or in junction with our concurrence result,
\begin{equation}
\braket{\mathcal{B}}=2\sqrt{2} \; C.
\end{equation}
This is consistent with the maximum expression of  $|\langle \mathcal{B} \rangle|_{\text{max}}$  since
\begin{equation}
\braket{\mathcal{B}}=2\sqrt{2} \; C < |\langle \mathcal{B} \rangle|_{\text{max}} = 2\sqrt{1 + C^2},
\end{equation}
as $0\leq C \leq 1$.
 
\section{Conclusions}
The quantum entanglement measure of concurrence has been shown to be directly calculable from a
Tomita-Takesaki modular operator framework constructed from the local von Neumann algebras
of observables for both a supersymmetric quantum mechanical system and a Bosonic quantum field interacting with an entangled pair of Unruh-DeWitt qubit style detectors. Specifically, the expectation value of the Tomita-Takesaki modular conjugation operator $J$ was used to find an expression for the
quantum concurrence $C$ of a bi-variate entangled state for both systems. As the modular conjugation operator $J$ links two separate
systems creating a form of duality with respect to their von Neumann algebras of observables, our
concurrence relation provides a direct physical meaning to this anti-unitary, anti-linear operator as both a symmetry operator and a quantitative measure of entanglement. The concurrence result for the Unruh-DeWitt detector model is shown to be in agreement with some known results on the Bell-CHSH inequality for such a system.

\bigskip

\section*{Appendix}

See \cite{KR1983, Tak1979, Bratteli1987} for extensive reviews of the topics outlined below.

\noindent\textit{\textbf{$\bm{C^*}$-Algebra:}} A $\bm{C^*}$-algebra $\mathcal{A}$ is a Banach algebra with an involutive map $ ^*: \mathcal{A} \rightarrow \mathcal{A} , \,\,\, A \rightarrow A^* \,\,\, \forall A \in \mathcal{A} , \lambda \in \mathbb{C}$ such that

\medskip

\noindent$(A^*)^*=A$

\noindent$(AB)^*=B^* A^*$

\noindent$(\lambda A)^*= \bar{\lambda} A^*$

\noindent$ (\lambda  A + \alpha B )^* = \bar{\lambda } A^* +\bar{\alpha}  B^*  $ (anti-linear)

\noindent$||A^*|| = ||A|| $

\noindent$||A A^*|| = ||A|| \, ||A^*|| = ||A^* A|| = ||A^*|| \, ||A|| =||A||^2$ (the $\bm{C^*}$ condition).

\bigskip

\noindent\textit{\textbf{Von Neumann Algebra:}} Consider a  $\bm{C^*}$-algebra $\mathcal{B(H)} =\{A\}$ of bounded linear operators on a Hilbert space, $ A: \mathcal{H} \rightarrow \mathcal{H}$. Let $\mathcal{C}$ be a subset of $\mathcal{B(H)}$. An operator $A \in \mathcal{B(H)} $ belongs to the commutant $\mathcal{C}'$ of the set $\mathcal{C}$ $\iff$   $AC =CA, \,\,\, \forall C \in \mathcal{C}$. A von Neumann algebra $\mathcal{A}$ is a unital $C^*$-subalgebra of $\mathcal{B(H)}$ such that $\mathcal{A}'' = \mathcal{A}$. Von Neumann proved that this bicommutant definition is equivalent to the algebra $\mathcal{A}$ being closed with respect to the weak topology on $\mathcal{B(H)}$. A von Neumann algebra in standard form is one where there exists an element $| \Omega \rangle \in \mathcal{H}$ which is both cyclic (operating on $| \Omega \rangle $ with elements in  $\mathcal{A}$ can generate a space dense in $\mathcal{H}$) and separating (if $A | \Omega \rangle = 0$, then $A=0$).

\bigskip

 \noindent\textit{\textbf{Tomita-Takesaki Modular Operators:}} Consider a  von Neumann algebra $\mathcal{A} \subset \mathcal{B(H)}$ in standard form with a cyclic and separating vector $| \Omega \rangle \in \mathcal{H}$. Let $S:\mathcal{H} \rightarrow \mathcal{H}$ be a anti-unitary operator defined by $S A  | \Omega \rangle = A^* | \Omega \rangle $. Let the closure of $S$ have a polar decomposition given by $S=J \Delta ^{\frac{1}{2}} =\Delta^{-\frac{1}{2}} J$, where $J$ is called the modular conjugation operator and $\Delta$ is called the modular operator. $J$ is anti-linear and anti-unitary whereas $\Delta$ is self-adjoint and positive. Furthermore, the following relations hold:

1. $J \Delta ^{\frac{1}{2}} J = \Delta ^{-\frac{1}{2}}$

2. $ J^2 =I, \,\,\, J^{*} = J$

3. $ J |\Omega \rangle  = |\Omega \rangle  $

4. $ J \mathcal{A} J = \mathcal{A}' $

5. $ \Delta = S^{*} S$

6. $ \Delta |\Omega \rangle  = |\Omega \rangle $

7. $ \Delta^{it} \mathcal{A} \Delta^{-it} = \mathcal{A} $ (one parameter-$t$ group of automorphisms of $\mathcal{A})$

8. If $ \omega (A) = \langle \Omega | A \Omega \rangle $, $ \forall A \in \mathcal{A}$,  then $\omega$ is a KMS (Kubo-Martin-Schwinger) functional (state) on $\mathcal{A}$ with respect to the automorphism of 7.

9.$|\Omega \rangle $  is cyclic for $\mathcal{A}$ if and only if $|\Omega \rangle $  is separating for $\mathcal{A}'.$


\begin{thebibliography}{10}

	  \bibitem{Chatterjee1}
		R. Chatterjee and T. Yu,
		\newblock {\em Modular Operators and Entanglement in Supersymmetric Quantum Mechanics},
		\newblock Journal of Physics A: Mathematical and Theoretical, 54 (20) (2021).

  	\bibitem{Chatterjee2}
		C. Gallaro and R. Chatterjee,
		\newblock {\em A Modular Operator Approach to Entanglement of Causally Closed Regions},
		\newblock International Journal of Theoretical Physics, 61 (8), (2022).

        \bibitem{Witt2022}
        E. Witten,
         \newblock {\em Why Does Quantum Field Theory in Curved Spacetime Make Sense? And What Happens to the Algebra of Observables in the Thermodynamic Limit?},
         \newblock{Dialogues Between Physics and Mathematics, Ed. Mo-Lin Ge and Yang-Hui He, Springer, New York, 2022.}
        
	  \bibitem{Witt2018}
		E. Witten,
		\newblock {\em Invited article on entanglement properties of quantum field theory},
		\newblock Review of Modern Physics, Volume 90, (2018).

        \bibitem{Sewell2002}
		G.L. Sewell,
		\newblock{\em Quantum Mechanics and its Emergent Macrophysics},
		\newblock Princeton University Press, 2002.

       \bibitem{Guedes2024}
        F.M. Guedes, M. S. Guimaraes, I. Roditi, and S. P. Sorella,
        \newblock {\em Unruh-De Witt detectors, Bell-CHSH inequality and Tomita-Takesaki theory},
        \newblock J. High Energy Phys. 2024, 31, (2024).

		\bibitem{Bratteli1997}
		O. Bratteli and D. Robinson,
		\newblock{\em Operator Algebras and Quantum Statistical Mechanics II, 2nd Ed.},
		\newblock Springer Verlag, Berlin, 1997.
  
  	\bibitem{Haag1992}
		R. Haag,
		\newblock{\em Local Quantum Physics, 2nd Ed.},
		\newblock Springer-Verlag, Berlin, 1996.


          \bibitem{summers1987I}
          S.J. Summers and R. Werner,
          \newblock{\em Bell’s inequalities and quantum field theory. I. General setting},
          \newblock J. Math. Phys., 28, 2440, (1987).

          \bibitem{summers1987II}
          S.J. Summers and R. Werner,
          \newblock{\em Bell’s inequalities and quantum field theory. II. Bell’s inequalities are maximally violated in the vacuum},
          \newblock J. Math. Phys., 28, 2448, (1987).

           \bibitem{Fabritiis2023}
          P. De Fabritiis, F.M. Guedes, M.S. Guimaraes, G. Peruzzo, I. Roditi, and S.P. Sorella,
          \newblock{\em Weyl operators, Tomita-Takesaki theory, and Bell-Clauser-Horne-Shimony-Holt inequality violations},
          \newblock Phys. Rev. D,  108, 8, (2023).

       \bibitem{Woot2001}
        W.K. Wooters,
        \newblock {\em Entanglement of Formation and Concurrence},
        \newblock Quantum Information and Computation 1, 1, (2001).

        \bibitem{Beng2017}
        I. Bengtsson and K. Zyczkowski,
        \newblock {\em Geometry of Quantum States, 2nd Ed.},
        \newblock Cambridge University Press, 2017.
        
        \bibitem{Ezawa2008}
        M. Ezawa,
        \newblock {\em Supersymmtric structure of quantum Hall effects in graphene},
        \newblock Phys. Lett. A. 372, 6, (2008).

        \bibitem{Uhlmann2000}
        A. Uhlmann, 
        \newblock {\em Fidelity and concurrence of conjugated states},
        \newblock Phys. Rev. A. 62, 3, (2000).
	
		\bibitem{Unruh1976}
		W. Unruh,
		\newblock {\em Notes on Black Hole Evaporation},
		\newblock Phys. Rev. D 14, 4, (1976).

        \bibitem{DeWitt1979}
        B. DeWitt,
        \newblock{\em Quantum Gravity: the new synthesis,}
        \newblock{General Relativity: An Einstein Centenary Survey, Ed. S.W. Hawking and W. Israel, Cambridge University Press, 1979.}
        
        \bibitem{Tjoa2023}
        E. Tjoa, 
        \newblock{\em Nonperturbative simple-generated interactions with a quantum field for arbitrary Gaussian states}
        \newblock Phys. Rev. D 108, 4, (2023).

	
		\bibitem{KR1983}
        R.V. Kadison and J.R. Ringrose,
        \newblock {\em Fundamentals of the Theory of Operator Algebras, Vol I},
        \newblock Academic Press, New York, 1983.

        \bibitem{Tak1979}
        M. Takesaki,
        \newblock {\em Theory of Operator Algebras I},
        \newblock Springer-Verlag, New York, 1979.

		\bibitem{Bratteli1987}
		O. Bratteli and D. Robinson,
		\newblock{\em Operator Algebras and Quantum Statistical Mechanics I, 2nd Ed.},
		\newblock Springer Verlag, Berlin, 1987.

		\end{thebibliography}
\end{document}